# An Integrated (Crop Model, Cloud and Big Data Analytic) Framework to support Agriculture Activity Monitoring System

Shamim Akhter
Computer Science and Engineering (CSE)
East West University (EWU), Dhaka, Bangladesh
shamimakhter@gmail.com

Kento Aida
Center for Grid Research and Development,
National Institute of Informatics (NII), Japan
aida@nii.ac.jp

Kiyoshi Honda
Chubu University
Aichi, Japan
hondak@isc.chubu.ac.jp

Amor V.M. Ines
IRI, Columbia University
New York, USA
ines@iri.columbia.edu

*Abstract*—Agriculture activity monitoring needs to deal with large amount of data originated from various organizations (weather station, agriculture repositories, field management, farm management, universities, etc.) and mass people. Therefore, a scalable environment with flexible information access, easy communication and real time collaboration from all types of computing devices, including mobile handheld devices as smart phones, PDAs and iPads, Geo-sensor devices, and etc. are essential. It is mandatory that the system must be accessible, scalable, and transparent from location, migration and resources. In addition, the framework should support modern information retrieval and management systems, unstructured information to structured information processing, task prioritization, task distribution, workflow and task scheduling system, processing power and data storage. Thus, High Scalability Computing (HSC) or Cloud based system with Big data analytics can be a prominent and convincing solution for this circumstance. In this paper, we are going to propose an integrated (crop model, cloud and big data analytics) geo-information framework to support agriculture activity monitoring system.

## I. INTRODUCTION

Agricultural activity monitoring, enclosed quantifying the irrigation scheduling, tracing the soil hydraulic properties, generating the crop calendar, prediction on crop growth in terms of planting date, acreage, planting intensity, water stress, biomass, yield etc., is important for efficient food security management at country level. It can also contribute to better policymaking, timely countermeasures, optimization of water resources distributions, damage assessment and finally to food supply security and stable market. Farmers want to know the above information in a regular basis. Researchers of agriculture also try to analyze various information about crops in order to take measures if they had some problems. Particularly, when an on-going experiment covers large area such as a country, Remote Sensing (RS) plays a vital role by providing useful information over large areas. However, some information, or crop parameters, e.g. ground water level, cropping season time extent, date of emergence of crop, irrigation scheduling cannot be visible directly through RS images, which reflects a practical problem that we cannot generate or observe those parameters from remote places. To collect those data, time by time basis field experiments are required. This is a time consuming, complex and expensive procedure. To overcome such problems, indirect methods such as inverse modeling with crop model can be used to obtain those basic input parameters. One such method is the manual calibration by "trial and error" procedure, which is very subjective and time consuming and uncertainty associated with them cannot be quantified. A more robust way of inverse modeling is to combine the model with optimization algorithm. However, processing the inverse modeling with crop model has a problem in practicality, that is, they require a huge amount of processing times. It is necessary to introduce methods for using higher processing power such as High Performance Computing (HPC) technologies. Some protocols or tools have been developed concerning the inverse modeling techniques and their HPC implementation models. However, the interoperability protocol between those agriculture applications and existing remote sensing (RS) image processing software is also necessary to improve practicality. On this regard, CAM-GA application has been imported with GRASS (Geographical Resource Allocation Software System) and implemented on single and multiple clusters environment. Different data and task distribution issues were experimented and analyzed. However, still the system is deficient regarding scalable environment issues- flexible information access, easy communication and real time collaboration from all types of movable computing devices, including mobile handheld devices as smart phones, PDAs and iPads, Geo-sensor devices, and etc. The system is also need to be accessible, scalable, and transparent from location, migration and resources. In addition, the framework should support modern information retrieval and management systems, unstructured information to structured information processing, task prioritization, task distribution, workflow and task scheduling system, processing power and data storage. Thus, High Scalability Computing (HSC) or cloud based system with big data analytics can be a prominent and convincing solution for this circumstance. In this paper, we are proposing an integrated (crop model, cloud and big data analytics) and service oriented geo-information framework to support agriculture activity monitoring system and discuss the implementation issues.

## II. MOTIVATION

### A. Inverse Modeling Techniques

Crop models, Soil-Water-Air-Plant (SWAP) [42] or Decision Support System for Agro technology Transfer (DSSAT) [41], have capacity to simulate soil, water and crop processes and serve as crop productivity monitoring tool. Crop Assimilation Model (CAM) predicts parameters of crop models with satellite images. A new methodology was developed in [27], called CAM-GA, to analyze the crop model (SWAP) parameters assimilation with Remote Sensing (RS) data and the assimilation procedure was optimized by an evolutionary searching technique called Genetic Algorithm (GA). CAM with double layers GA, CAM-DLGA [9], uses directly visible multi-resolution RS images [10] [29] and inversely assimilates to SWAP model data for estimating the non-visible model parameters. Other similar functionality models, e.g., CAM-PSO [28] and CAM-PEST [16], use different evolutionary searching techniques. However, processing the agricultural information with CAM has a problem in practicality that is, they require a huge amount of processing times. It becomes necessary to introduce methods for using higher processing power such as High Performance Computing (HPC) technologies.

### B. High Performance Computing Issues

Multi-computer based distributed systems (Clusters and Grids) have a large processing capacity for a lower cost; naturally, choice turns towards developing HPC applications. However, it was not an easy job to port CAM on HPC environment. The application performance was significantly affected by the data and task distribution methods on HPC. The workload in HPC, the bandwidth, the processors speed, parameters of evaluation methods and data size were additional concerning factors. The developers of agriculture or satellite image processing applications required solving the problem of both data and task distribution issues, or how to distribute data and tasks among single or multiple clusters environment. CAM-GA Model was successfully implemented on cluster computers and the implementation strategies are described in [4]. Results and experiments illustrate that the distributed pixel model performs well in cluster computing testbed [2][5]. Interoperability between the agriculture application and existing satellite image processing software is also necessary to improve practicality.

GRASS GIS is an open source software/tool, which is used to process satellite images. Inside GRASS, different modules have been developed for processing satellite images. A framework was developed to enable the GRASS GIS environment for satellite image processing on distributed computing systems, as an example basis [6]. Afterward, the CAM-GA application was imported as a GRASS module. High demand of distributing behavior was appealed inside the GRASS CAM-GA modules. The performance improvement of CAM is discussed in [3] and the work includes parallel methods of CAM-GA. However, the discussion in [3] is limited to parallel approaches only, and their performance analysis is missing. In particular, the performance is significantly affected by task distribution methods on the Grid, and developers of RS applications need to solve the problem of task distribution, or how to distribute tasks among multiple clusters on the Grid. Three different CAM-GA parallel methods, the pixel distribution, the population distribution, and the hierarchical distribution were successfully implemented and their performances were compared through the experiments on the real Grid testbed. These methods employed GridRPC as the programming framework but ways of task distribution were different [5]. The results in [5] demonstrate that the pixel distribution model exhibits the lowest communication overhead and presents the best performance on some setting of Grid. However, the hierarchical distribution model is the most preferable model to employ more parallelism in the CAM-GA application [8]. Afterwards, the same research was extended with real experimental data and implemented on real Grid testbed [7].

### C. Web Portal Implementation Issues

Web based technology can contribute an understandable, efficient and effective agriculture activity monitoring system. It is hard for agricultural researchers to understand the HPC systems and to port their agriculture models on them. Some researches separately focused on the performance improvement of agricultural model activities with HPC [31], [11], [12] or satellite image processing with HPC [17], [18], [15], [39], [53], [35], [40] and very few of them are in the merging domains on satellite image processing, agriculture and HPC. Implementing the GRASS modules and visualizing output images from web, is not a new concept. An excellent, fast and flexible open source web mapping software is UMN/Mapserver [20]. WMS (Web Map Service), is an online access to, or integration and exploitation of geospatial information, was successfully implemented [32] and [38] with Free Open Source Software (FOSS) to provide satellite image server. GRASSLink [26], a UNIX shell script based model, and it is capable to access spatial information from GRASS datasets and displays necessary parameters in web through web-mapping applications. A utilization of DHTML and JavaScript technology also enhanced more interaction from GRASSLinks application as developed in [2001]. PyWPS, Python Web Processing Service, was introduced in April, 2006 [36] and provides an implementation on Open Geospatial Consortium (OGC)'s web processing service standards. The realistic structure of the distributed agriculture monitoring system with a web portal on HPC and its implementation for larger domain like country or province is proposed and discussed in Figure 2 with the following schemes:

- processing satellite images automatically through HPC,
- CAM HPC implementation with appropriate data and task distribution schemes.

Individually, each step in Figure 1 is discussed and implemented with different data and application domains in [6] and [7]. GRASS GIS [2009] tool is used to process the satellite images automatically. CAM is implemented as a GRASS module. The interconnectivity between the GRASS on HPC platform is successfully established and then the GRASS CAM module is implemented in HPC platform with different data and task distribution methods. However, their combined framework for modeling the overall distributed agriculture monitoring scheme is not yet been established. Thus, agriculture researchers require a web based system or tool for agriculture activity monitoring so that they do not need to concern about the implementation issues for agricultural models or RS image processing on HPC.

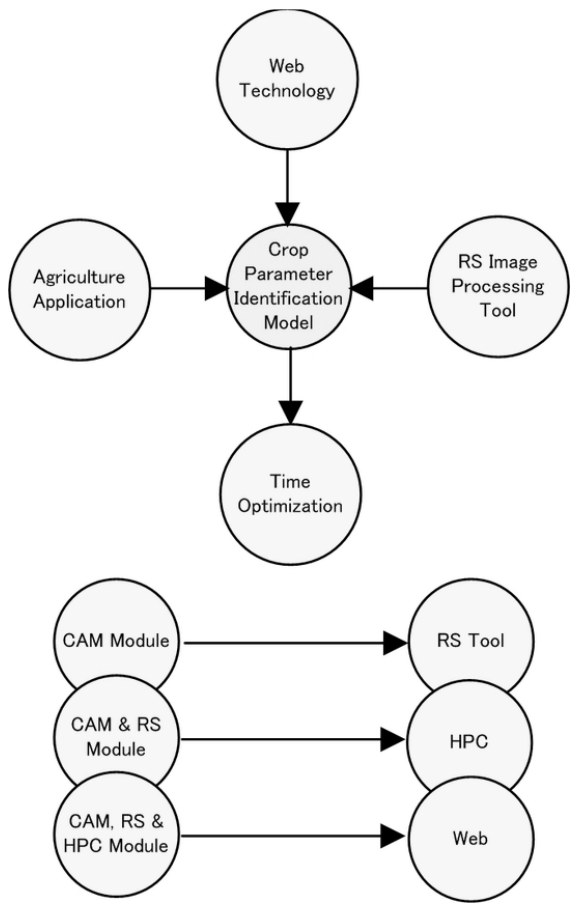

Figure 1: Essential Components to Build the Framework

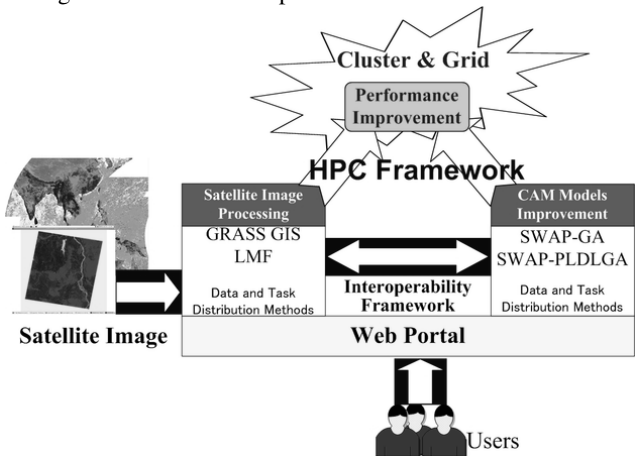

Figure 2: The Realistic Agriculture Framework with Web Portal and HPC

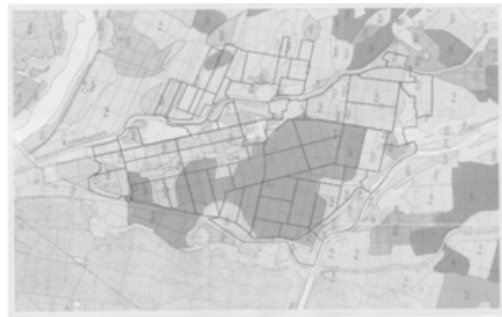

Figure 3: The Soil Map File

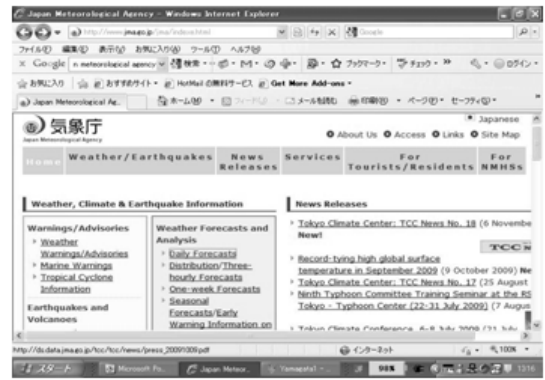

Figure 4: The Japan Metrological Agencies web site

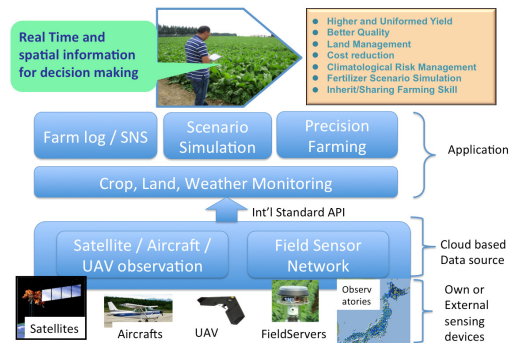

Figure 5: Overview of FieldTouch Agricultural Information Service Platform

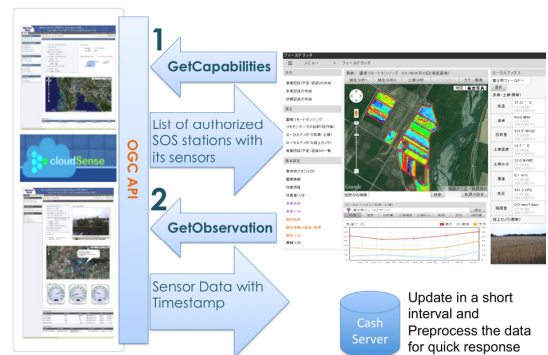

Figure 6: Sensor Observation Service

*D. Data Collection and Integration Issues*

Data collection, integration and processing appropriate information from data repositories are additional concerns for this research implementation. This research will co-ordinate with large amounts of data. Figure 3 provides an example of soil map, and the figure 4 provides Japan Metrological Agencies web site to collect metrological data for crop model. FieldTouch [25], an agri-information service platform (Figure 5), is built and tested on geospatial data infrastructure and CAM frameworks. FieldTouch integrates multi-scale sensor data for field monitoring, provides functionality for recording agricultural practices and then supports farmers in decision making. Field sensor network works to collect real-time data from distributed sensors and provides the data via Internet [23]. Sensor background service called cloudSense [24] works to ensure easy connection of field sensors to database and interoperability based on SOS (Sensor Observation Service) (Figure 6). These kinds of works together integrate into a single data repository. However, country level agriculture monitoring requires hundreds to thousands data repositories and those data repositories frequently update their own resources. Thus, interoperability, reliability and accessibility between the repositories are also important issues.

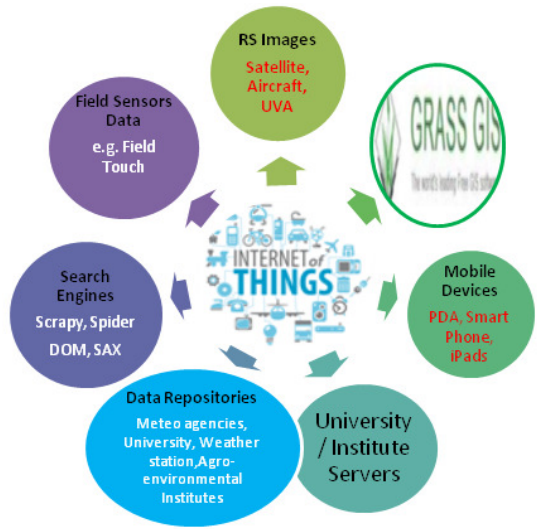

Figure 7: Data Sources

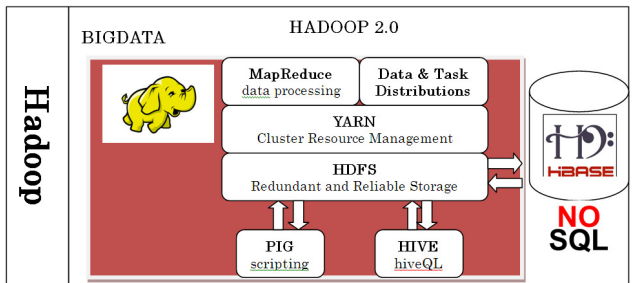

Figure 8: Hadoop Big Data Framework

## III. PROPOSED FRAMEWORK: INTEGRATION OF CAM, BIG DATA AND CLOUD ARCHITECTURE

Based on the above discussion, large scale agricultural activity monitoring requires to congregate information from Remote Sensing (RS) images and that type of processing takes a huge amount of computational time and optimization on the computational time is a vital requirement. High Performance Computing (HPC and parallel computing) can be a solution but not enough. Agriculture activity monitoring also needs to deal with large amount of data originated from various organizations (weather station, agriculture repositories, field management, farm management, universities, agencies etc.) and mass people. Therefore, a scalable environment with flexible information access, easy communication and real time collaboration from all types of computing devices, including mobile handheld devices (smart phones, PDAs and iPads and etc.), Geo-sensor devices (spectrometer and turbidity sensors, and etc.) are essential. It is mandatory that the system must be accessible, scalable, and transparent from location, migration and resources. In addition, the framework should support modern information retrieval and management systems, unstructured information to structured information processing (IBM Info Stream [43], text analytics, pig [45] &/ hive [46] etc.), task prioritization, task distribution (Hadoop [44]), workflow and task scheduling system[21][22], processing power and data storage (Amazon S3, Google BigTable [34] and Google BigQuery [47]. Thus, High Scalability Computing (HSC) or Cloud based system can be a prominent and convincing solution for this circumstance.

Data sources (Figure 7) are the RS images, GRASS Database resources, mobile handheld devices, such as smart phones, PDAs and iPads, Geo-sensor devices, such as spectrometer and turbidity sensors, field sensors (Field Touch), Data repositories (Meteo agencies, University, Weather station, Agro-environmental Institutes), University/Institution servers etc. Data sources frequently update their own resources, and interoperability of different kind of data with their different formats is also necessary. Thus, stream computing techniques or search engines can intelligently crawl data repositories with meta data indexing (title, description, keyword etc.) to capture real time unstructured data streams, convert them to the required structure (xml parsing), and store in database. Later, structured data can be queried and analyzed to sort, group, and filter in order to answer agro-business questions or measure framework capabilities. Additional tools are required to efficiently collect, effectively aggregate, and move large amounts data from the sources to big data framework (Figure 8). Flume [50] or Sqoop [51] can provide flexible architecture based on streaming data flows to support robust and fault tolerant mechanism for transferring data from sources to big data framework. The Big data framework consists open-source Hadoop 2.0 (clusters) for storing and analyzing massive amounts of structured and unstructured data with YARN (Yet Another Resource Negotiator) and MapReduce based distributed processing framework. Parallel CAM-GA [8] can carry its own distributions (data & task) or share MapReduce for computation, distribution or processing. To store such huge data, the files are distributed over multiple machines/nodes. Hadoop Distributed File System (HDFS)-a java based file

Table1: Interplay Relationship between Cloud and Big Data

|      | **Volume** | **Variety** | **Velocity** |
|------|-----------|-------------|--------------|
| SaaS | Data & Semantics | Visualization | Real-time |
| PaaS | Distributed Processing | Schemaless | Integration on the fly |
| IaaS | Scalable storage | Federated store | On-demand resources e.g Cluster size, local/Cloud storage |

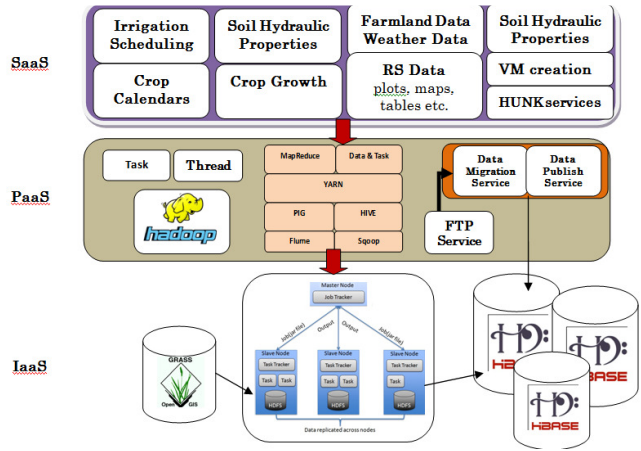

Figure 9: Proposed Cloud based Agriculture Framework

system-provides scalable, reliable and redundant data storage with high fault tolerant support. HDFS is good for sequential data access and lacks of random read/write option. Pig is a scripting language for creating MapReduce and data analysis programs on Hadoop. Since Pig is amenable to substantial parallel operations, thus enables to process enormous amounts of data very easily and quickly. Hive is a data warehouse system built for Hadoop and allows easy data aggregation, ad-hoc queries and analysis of large data sets stored in HDFS. HiveQL is a SQL "like" language that can be used to interact with the data and allows developers to put in their own custom mappers/reducers. HBase (Hadoop Base) is a NoSQL, distributed, scalable, big data store and able to perform real-time read/write access. It provides a fault-tolerant way of storing large quantities of sparse data pre-processed using Pig&/Hive. Hunk [48] is an analytical tool to analyze and visualize data in Hadoop. CAM model follows Master-Worker job distribution paradigm and thus can easily be implemented on Hadoop environment. CAM model can access HBase for required input data and processed using MapReduce computation.

Hadoop is still facing challenges due to its implementation limitations including- need to have all the tools before running the job, cost to buy hardware, and prior knowledge of processing is required to choose appropriate/optimum number of cluster nodes/resources to run the job. It also requires proper planning to tradeoff between cost and resources, specialized expertise is needed to use Hadoop effectively. Cloud eliminates the operational challenges of running Hadoop and provides- elastic platform &/ service (expand/shrink cluster size) as data and processing requirements, no hardware cost, do not need to pay in advance- pay what is needed, and easy deployment facility.

Big data supports volume, variety and velocity whereas Cloud supports three Layers of services: Software as a Service (SaaS), Platform as a Service (PaaS), and Infrastructure as a Service (IaaS). In [52], Tony has drawn an interplay relationship (Table 1) between Cloud and Big Data. Architecture of private Cloud for agriculture framework with big data analytics are presented in Figure 9. Remote Sensing image processing tool ArcGIS [1] is already implemented in Cloud. GRASS is also implemented on Cloud environment. However, the three tiers architecture, GRASS-Hadoop-Cloud, will provide a new dimension of research for GRASS/Geo Science community. For the web implementation part, Openlayer [33] or PyWPS [36] both are able to support dynamic mapping concept with interconnection between the image repositories with necessary security infrastructure.

PyWPS is relatively new concept. It is an implementation of the Web Processing Service standard from the Open Geospatial Consortium. Connection establishment between PyWPS and GRASS is relatively easier. Additionally, it supports dynamic GRASS location creation for a given input image during the execution phase, makes the task easier for sharing input images from another repository. The web portal will be an integration of PyWPS, Hunk and Ambari. Hunk supports Hadoop as search engine and processes data using MapReduce. Ambari [49] provides an intuitive, easy-to-use Hadoop monitor and management web interface. CAM web portal implementation needs two separate modules for client and server. User can select the specific region from a given image through the interface. Hunk will trace the user given queries and submit those queries to the server modules. Server module will execute the GRASS based CAM module with user given queries and generate unknown crop information; those are not directly extractable from RS images.

IV. CONCLUSION

Cloud based system will provide a scalable environment with lower cost, offsite/onsite data storage, flexible information access, easy communication and real time collaboration between RS image and agriculture data repositories. The system will be accessible, scalable, and transparent from location, migration and resources. Real time data collection and integration with intelligent system (data frame and algorithm) supports can provide successful automation and data storage. In future those data can be used for any further prediction purpose. The successful implementation of this kind research helps the policy makers to monitor the on field agriculture behavior and takes prompt decision and action regarding any unusual condition. Through the Cloud based web portal (on Hadoop infrastructure) the user can select the specific area for running the agriculture model and later the outputs will be generated regarding the specific crop parameters which are not directly visible from RS image. Users don't need to bother about the backend processes like the crop models or HSC or the distribution mechanisms. The expected time for processing the user request will be within hour range. Multi-users can interact in the system at the same time. So, the portal must be capable to serve multi users processing requests.

## REFERENCES

[1] ArcGIS, 2012, ArcGIS as a System for Emergency/Disaster Management, ArcGIS as a System for Emergency/Disaster Management, Last visited 11.11.2012

[2] Akhter, S., Honda, K., Chemin, Y., Uthayopas, P., 2005, Distributed Pixel Method to speed-up the RS data assimilation of SWAP model, In Proceedings of the MapAsia conference, Jakarta, Indonesia.

[3] Akhter, S., Jangjaimon, I., Chemin, Y., Uthayopas, P., and Honda, K., 2006, Development of a GRIDRPC tool for Satellite Images Parallel Data Assimilation in Agricultural Monitoring, International Journal of Geoinformatics,ISSN 1686-6576, Vol.2 No.3.

[4] Akhter, S., Honda, K., Chemin, Y., and Uthayopas, P., 2007a, Exploring Strategies for Parallel Computing of RS Data Assimilation with SWAP-GA , Journal of Computer Science, 3(1): pp. 47-50, 2007.

[5] Akhter, S.,Osawa, K., Aida, K., 2007b, Performance Evaluation of Distributed SWAP-GA Models with GridRPC, IPSJ SIG Technical Reports, 2007-HPC-111, pp.133-138, Aug. 2007.

[6] Akhter, S., Chemin,Y., Aida, K., 2007c, Porting a GRASS raster module to distributed computing: Examples for MPI and Ninf-G, *OSGeo Journal*, Vol2., pp.36-44.

[7] Akhter, S., Osawa, K., Nishimura, M., and Aida,K., 2008, Experimental Study of Distributed SWAP-GA Models on the Grid, *IPSJ Transactions on Advanced Computing Systems*, Vol.1 No.2, pp.193-206.

[8] Akhter, S., Aida, K., Chemin, Y., 2010a, GRASS GIS on High Performance Computing with MPI, OpenMP and Ninf-G Programming Framework, Proceeding of ISPRS 2010, 9-12 Aug Kyoto, Japan.

[9] Akhter, S.,Sakamoto, K., Chemin, Y., Aida, K., 2010b, Self-organizing GA for Crop Model ParameterEstimation using Multi-resolution Satellite Images, International Journal of GeoInformatics, Vol. 6, No. 4, Dec, 2010, ISSN 1686-6576.

[10] ASTER Image Webpage, 2009, http://asterweb.jpl.nasa.gov/, Internet.

[11] Aqeel-ur-Rahman and Shaikh, Z.A. (2008), Towards Design of Context-Aware Sensor Grid Framework for agriculture, Fifth International Conference on Information Technology (ICIT), XXVIII-WASET Conference, April 25-27 Rome, Italy.

[12] CALS, 2007, Cornell's College of Agriculture and Life Sciences, Cornell's Supercomputer will Crunch Weather Data to Help Farmers Manage Chemicals, http://www.news.cornell.edu/stories/May07/HPCinAg.ws.html, Internet.

[13] Carlson, T.N., Taconet, O.,Vidal, A., Gilles, R.R., Olioso, A., and Humes, K. (1995), An Overview of the Workshop on Thermal Remote Sensing Held at La-Londe-Les-Maures, France, September 20-24, 1993,Agriculture and Forest Meteorology, 77(3-4), 141-151.

[14] Condit, H. R. (1970), The Spectral Reflectance of American Soils, Photogrammetric Engineering, 36(9), 955-966.

[15] Dong,S. and Hu,Q. (2005), Building Remote Sensing Database on Grid, Geoscience and Remote Sensing Symposium, IGARSS '05 Proceedings, IEEE International, 25-29.

[16] Dorji, M., 2003, Integration of SWAP Model and SEBAL for Evaluation of on Farm, Irrigation Scheduling with Minimum Field Data, Enschede, ITC, 100 p.

[17] GEO Grid (2005), http://www.geogrid.org/, Internet.

[18] GOEN (The Geosciences Network) (2002), http://www.geongrid.org/, Internet.

[19] GRASS GIS, 2009, Geographic Resources Analysis Support System, http://grass.itc.it/, Internet.

[20] GRASSMAP, 2009, Simple demonstrational GRASS/UMN/MapServer( Spearfish data) (2009), http://grass.itc.it/start.html ,Internet.

[21] Habiba,M., Akhter, S., 2012, MAS Workflow Model and Scheduling Algorithm for Disaster Management System, International Conference On Cloud Computing Technologies, Applications And Management (ICCCTAM-12), UAE-Dec 8-10,2012.

[22] Habiba, M., Akhter, S., 2013, A Cloud Based Natural Disaster Management System. The 8th International Conference on Grid and Pervasive Computing (GPC 2013), Seoul Olympic Parktel, Seoul, Korea, May 9-11, 2013.

[23] Hirafuji, M., Fukatsu, T., Hu, H., Yoichi, H., Kiura, T., Ninomiya, S. Wada, M. Shimamura, H., 2005, Field Server: Multi-functional Wireless Sensor Network Node for Earth Observation, pp.304, SenSys'05 Proceedings of the Third International Conference on Embedded Networked Sensor Systems, November 2-4, 2005, San Diego, California, USA

[24] Honda, K., Shrestha, A., Witayangkurn, A., Chinnachodteeranun, R. and Shimamura, H., 2009, Fieldservers and Sensor Service Grid as Real-time Monitoring Infrastructure for Ubiquitous Sensor Networks. Sensors (ISSN 1424-8220), 9, no. 4: 2363-2370., doi:10.3390/s90402363, 2009

[25] Honda, K., Ines, A.V.M., Yui,A., Witayangkurn, A., Chinnachodteeranun, R. and Teeravech, K., 2014, Agriculture Information Service Built on Geospatial Data Infrastructure and Crop Modeling IWWISS '14, Sep 01-02 2014, Saint Etienne, France, ACM 978-1-4503-2747-3/14/09. http://dx.doi.org/10.1145/2637064.2637094

[26] Huse, S.M. (1995), GRASSLinks: A New Model for Spatial Information Access for Environmental Planning, PhD Thesis, University of California, URL: http://www.regis.berkeley.edu/sue/phd/.

[27] Ines, A.V.M., 2004, Improved Crop Production Integrating GIS and Genetic Algorithms, *PhD Thesis,* Asian Institute of Technology (AIT), KhlongLuang, Bangkok, Thailand, AIT Diss No.WM-02-01.

[28] Kamble, B., Chemin, Y.H., 2006, GIPE in GRASS Raster Add-ons, http://grass.gdf-hannover.de/wiki/, GRASSAddOns, RasterAdd-ons, Internet.

[29] MODIS Image Webpage, 2009, http://modis.gsfc.nasa.gov/, Internet.

[30] Nazrov,E., 2011, Emergency Response management in Japan, Final Research report, ASIANDisaster Reduction Center, FY2011A Program.

[31] Ninomiya,S., Laurenson,M. and Kiura,T.(2009), Network Computing for Agricultural Information Systems-GRID for Agricultural Decision Support, http://www.google.com/searchhl=en\&q=Network+COmputing+for+ Agricultural +Information+Systems, Internet.

[32] Ninsawat, S. and Honda, K.(2004), Development of NOAA and Landsat Image Server using FOSS, Proceedings of the FOSS/GRASS Users Conference, Bangkok, Thailand, 12-14 September 2004.

[33] OpenLayer, 2009, OpenLayers: Free Maps for the Web, http://openlayers.org/, Internet.

[34] Pandey,S., Karunamoorthy, D., and Buyya,R., 2011, Workflow Engine for Clouds,CloudComputing:Principles and Paradigms, R. Buyya, J. Broberg, A.Goscinski (eds),ISBN-13: 978-0470887998, Wiley Press, New York, USA.

[35] Petrie,G. M., Dippold, C., Fann,G. , Jones,D., Jurrus,E., Moon,B. and Perrine K.(2002), Distributed Computing Approach for Remote Sensing Data, Proceedings of the 34$^{th}$ Symposium on. the Interface, April 17-20, Montreal, Quebec, Canada.

[36] PyWPS, 2009, Python Web Processing Service, http://pywps.wald.intevation.org/, Internet.

[37] Raghavan,V., Herath,S., Dutta,D. (2001), An Internet based Water Infrastructure Inventory System, International Symposium on Achievements of IHP-V in Hydrological Research, Hanoi, Vietnam, 345-351.

[38] Raghavan,V., Santitamnont, P., Masumoto,S. and Nemoto,T.(2004), Implementation of Web Map Server Test-bed and Development of Training Material for Advancing FOSS4G Solutions, Proceedings of the

FOSS/GRASS Users Conference, Bangkok, Thailand, 12-14 September 2004.

[39] Rokos, D.Kl. and Armstrong, M.P.(1998), Experiments in the identification and extraction of terrain features using a PC-based parallel computer, Photogrammetric Engineering and remote Sensing, 64(2), 135-142.

[40] Shen,Z., Luo,J., Huang,G., Ming,D.,Ma,W. and Sheng,H.(2007), Distributed Computing Model for Processing Remotely Sensed Images based on Grid Computing, Information Sciences, Volume 177, Issue 2,15, 504-518.

[41] Tsuji, G.Y., Uehara, G., and Salas, S., 1994, DSSAT v3.0., Honolulu, Hawaii: University of Hawaii.

[42] Van Dam, J.C., et. al., 1997, SWAP Model, http://www.swap.alterra.nl/, Internet. Web1, 2014, InfoSphere Streams, http://www.ibm.com/developerworks/bigdata/streams/

[43] Web2, 2014 Apache™ Hadoop, http://hadoop.apache.org/

[44] Web3, 2014 Pig, https://pig.apache.org/

[45] Web4, 2014 Hive, http://hive.apache.org/

[46] Web5, 2014 Google BigQuery, https://developers.google.com/bigquery/

[47] Web6, 2016 Hunk, http://www.splunk.com/en_us/products/hunk.html

[48] Web7, 2016 Ambari, http://hortonworks.com/hadoop/ambari/

[49] Web8, 2016 Flume, https://flume.apache.org/

[50] Web9, 2016 Sqoop, http://sqoop.apache.org/

[51] Web10, 2016 Big Data Cloudified, http://tonyshan.sys-con.com/node/2707730

[52] Yang,C.T.,Chang,C.L., Hung,C.C. and Wu,F.(2001), Using a Beowulf Cluster for a Remote Sensing Application, 22nd Asian Conference on Remote Sensing Proceeding, Singapore, 5-9 November, 2001.